# Orbital-resolved imaging of coherent femtosecond exciton dynamics in coupled molecules

Yang Luo[1,2]†, Shaoxiang Sheng[1,3]†, Michele Pisarra[4,5]†, Caiyun Chen[1], Fernando Martin[6,7]*, Klaus Kern[1,8], Manish Garg[1,*]

[1]Max Planck Institute for Solid State Research, Heisenbergstr. 1, 70569 Stuttgart, Germany

[2]Hefei National Laboratory, University of Science and Technology of China, Hefei, 230088, China

[3]Tsientang Institute for Advanced Study, Hangzhou, 310024, China

[4]Dipartimento di Fisica, Università della Calabria, Via P. Bucci, cubo 30C, 87036, Rende (CS), Italy

[5]INFN-LNF, Gruppo Collegato di Cosenza, Via P. Bucci, cubo 31C, 87036, Rende (CS), Italy

[6]Instituto Madrileño de Estudios Avanzados en Nanociencia (IMDEA Nano), Faraday 9, Cantoblanco, 28049 Madrid, Spain

[7]Departamento de Química, Módulo 13, Universidad Autónoma de Madrid, 28049 Madrid, Spain

[8]Institut de Physique, Ecole Polytechnique Fédérale de Lausanne, 1015 Lausanne, Switzerland

*Corresponding authors. Email: mgarg@fkf.mpg.de and fernando.martin@imdea.org

†These authors contributed equally to this work: Y. Luo, S. Sheng and M. Pisarra

**Abstract**

Optical excitation and control of excitonic wavepackets in organic molecules is the basis to energy conversion processes. To gain insights into such processes, it is essential to establish the relationship between the coherence timescales of excitons with the electronic inhomogeneity in the molecules, as well as the influence of intermolecular interactions on exciton dynamics. Here, we demonstrate orbital-resolved imaging of optically induced coherent exciton dynamics in single copper napthalocyanine (CuNc) molecules, and selective coherent excitation of dark and bright triplet excitons in coupled molecular dimers. Ultrafast photon-induced tunneling current enabled atomic-scale imaging and control of the excitons in resonantly excited molecules by employing excitonic wavepacket interferometry. Our results reveal an ultrafast exciton coherence time of ~ 70 fs in a single molecule, which decreases for the triplet excitons in interacting molecules.

**Main Text**

Molecular exciton dynamics is at the heart of numerous ultrafast processes, such as electronic energy transfer (*1, 2*) and charge separation in photosynthetic light-harvesting complexes (*3, 4*), and underpins the foundation of emerging quantum technologies (*5*). Owing to the atomic-scale variation of the valence electron density in molecules, excitons are generated with varying efficiency, and their coherence timescales and dynamics are influenced by the intermolecular interactions (*6, 7*). Such interactions also lead to the emergence of new quantum states (*8*). Coherent dynamics of the excitonic wavefunctions has been tracked in real-time in the bulk phase using wavepacket interferometry, where the relative phases of the involved eigenstates in the quantum superposition can be manipulated to yield the desired wavefunction (*9-12*). Nevertheless, experiments in the bulk provide an ensemble-averaged perspective, where both the atomic-scale quantum dynamic properties of the excitons and intermolecular interactions are smeared out (*13-15*). Moreover, the optically dark excitonic states are difficult to investigate in the far-field measurements (*7, 16, 17*).

The unification of scanning tunneling microscopy (STM) with ultrashort light pulses (*18-27*) provides a unique avenue to directly image ultrafast exciton dynamics in single molecules at their natural length (angstrom) and time (femtoseconds) scales. In the current work, we utilize such a quantum microscope coupled with a sequence of two ultrashort pulses to track excitonic wavefunctions in single molecules by wavepacket interferometry. The generation of atomically localized photocurrents in single molecules enabled orbital-resolved imaging and investigation of excitonic coherences. Excitons are efficiently generated only in the regions of high valence electron density and their coherence timescales are invariant to the local chemical functionality. In coupled molecule dimers, bright and dark triplet excitonic states emerge, which were selectively excited and probed at the atomic scale, determined by the position of the nanotip of the STM over the molecules. The excitonic coherence times in coupled molecules are shorter compared to the isolated single molecules, while the dark and bright excitonic states exhibit different coherence timescales.

**Femtosecond orbital-resolved photocurrent generation in a single molecule**

In our experiments, ultrashort laser pulses ($\tau \sim 15$ fs) illuminate single CuNc molecules present in the plasmonic tunnel junction of an STM comprising of a gold tip and a silver (Ag(111)) substrate, as schematically shown in Fig. 1A. An ultrathin dielectric film of NaCl beneath the molecules serves as an electronic decoupling spacer, isolating them from the underlying metallic substrate. STM topographic images of the single CuNc molecules on three monolayers (ML) of NaCl exhibits a four-lobe structure (Fig. 1B). Differential conductance measurement with the nanotip of the STM placed over the molecular lobe reveals two discrete peaks corresponding to the highest occupied and lowest unoccupied molecular orbitals, HOMO and LUMO, respectively (Fig. 1C). STM topographic images of the respective frontier orbitals of the CuNc molecule are shown in the insets of Fig. 1C. The spectral bandwidth of the illuminating laser pulses encompasses the optical gap of CuNc, as evident by the comparison of the STM-induced luminescence (STML) spectrum (red curve) of a single CuNc molecule (*8, 28, 29*) with the spectrum of the laser pulses (blue curve) in Fig. 1D. This spectral overlap allows for the direct excitation of excitons and the creation of coherent superposition of electronic states in the CuNc molecule.

Figure 1E compares the variation of the tunneling current as a function of the applied bias (*I-V* curve) in the tunnel junction acquired with ("Femtosecond photocurrent") and without ("DC current") laser pulse illumination. The *I-V* curves were measured in the bias range from −1 V to 0 V, chosen to be within the HOMO-LUMO tunneling gap (dashed rectangle in Fig. 1C), in order to avoid any DC tunneling current via the molecular orbitals. In the absence of laser illumination, the tunneling current is negligible at 0 V, and reaches ~ 100 fA at −1 V (black curve). In contrast, under laser pulse illumination, a substantial tunneling current emerges and reverses polarity around −300 mV when the bias is swept from −1 V to 0 V, which transparently indicates the contribution of the laser-induced tunneling current in the *I-V* curve. The reversal in the polarity of the laser-induced tunneling current suggests that the electrons can flow either from the tip to the molecule or vice versa, depending on the applied bias and the transient superposition of electronic states created by femtosecond laser pulses (*30, 31*).

Figure 1F shows the variation of the laser-induced tunneling current as a function of the laser power at two different biases of 0 V and −1 V, which display opposite polarities. An apparent linear scaling of the laser-induced tunneling current implies a weak-coupling regime of interaction (*32, 33*). The laser-induced tunneling current, hereafter referred to as photocurrent, arises solely from the absorption of a single photon by the molecules. A photocurrent of 100 fA generated by the laser pulses at a repetition rate of 80 MHz implies tunneling of mere ~ 0.008 electrons per laser pulse. This ultra-low tunneling yield per pulse highlights the quantum-limited nature of the single-photon excitation process and the sensitivity of our approach in probing single photon-driven exciton dynamics at the atomic scale.

The spatial extent of localization of the photocurrent was investigated by recording its variation on gradual change of the vertical position of the nanotip. An exponential dependence of the photocurrent was measured, as shown in Fig. 1G. The photocurrent vanishes on a relative increase of the nanotip height by ~ 150 pm. The sharp exponential decay further illustrates the extreme spatial confinement of the photocurrent at its atomic-scale origin.

Upon excitation with ultrashort laser pulses, the transient change of population in the electronic states gives rise to a highly localized tunneling current through the molecule, which allows for real-space imaging of the photocurrents with a very high spatial resolution. Figure 2A shows the photocurrent maps of a single CuNc molecule recorded at various bias voltages between 0 V and −1 V in the constant-height mode (open feedback) of the STM. Strikingly, the photocurrent images exhibit distinct patterns depending on the applied bias. At 0 V, a spatial distribution of positive photocurrent (red color) shows an 8-lobe pattern, whereas at −1 V, a spatial distribution of negative photocurrent (blue color) reveals a four-lobe feature. A gradual evolution in the spatial pattern and a reversal in photocurrent polarity was observed when the bias is swept from 0 V to −1 V.

To determine the relation between the recorded photocurrent images and the molecular orbitals, we recorded the standard DC tunneling current images of the orbitals (in the constant-height mode) without laser illumination as shown in Fig. 2B. The spatial patterns resembling the LUMO and HOMO of CuNc were measured at 1V and −2V, respectively, consistent with the energy alignment of the orbitals as revealed by the differential conductance measurement (Fig. 1C). Nonetheless, no clear orbital patterns were observed in the DC current images at 0 V or −1 V. The cross feature in the DC current image measured at −1 V instead corresponds to the geometrical structure of the molecule. A comparison of the photocurrent images with the DC current images indicates that the photocurrent image at 0 V resembles the orbital shape of the HOMO, whereas the photocurrent image at −1 V resembles the LUMO pattern. This observation clearly shows that the ultrafast photocurrent arises from transient photo-induced changes in the population of involved electronic levels, which projects the spatial character of HOMO or LUMO onto the tunneling current depending on the applied bias.

Figure 2C shows a schematic description of the plausible mechanism for the generation of photon-induced tunneling current at 0 V and −1 V, respectively. Ensuing excitation with a single photon (step 1), CuNc is excited to the $^2D_1$ (doublet) electronic state, resulting in the presence of an electron in the LUMO and one hole in the HOMO, i.e. a bound exciton. At a bias voltage of 0 V, electron transfer from the nanotip into the HOMO of the photoexcited molecule forms a transient anion (step 2), and subsequently, the electron in the LUMO tunnels to the substrate (step 3), yielding a net positive current. Since the spatial distribution of this photocurrent relies on electron injection from the nanotip to the molecule in step 2, the resulting photocurrent image manifests the spatial profile of the HOMO. In contrast, when the STM junction is biased at −1 V, the excited electron in the LUMO of CuNc tunnels to the nanotip, leaving a transient cation that is subsequently re-filled by an electron from the underlying Ag(111) substrate (Fig. 2D). Hence, this process yields a net negative photocurrent that reflects the spatial distribution of the LUMO. The second mechanism of the photocurrent generation (Fig. 2D) comes into play when the LUMO of the excited molecule lies above the Fermi level of the nanotip, i.e., when the bias voltage is lower than −0.3 V. This explains the change in the polarity of the photocurrent at approximately −0.3 V. The efficiency of the photocurrent generation around the center of the CuNc molecule is negligible (Fig. 2A), as the local dipole moment induced by the laser pulses cancels out due to the symmetry of the molecule.

We note that the mechanism of photocurrent generation in single molecules on excitation with resonant and off-resonant continuous wave (CW) laser sources have been discussed in recent works from other groups (*30, 31, 34*). In the current work, we have demonstrated orbital-resolved photocurrent generation in single molecules on illumination with ultrashort laser pulses, which is the key to imaging electron dynamics in single molecules.

**Orbital-resolved excitonic wavepacket interferometry**

Upon ultrafast photoexcitation of a CuNc molecule with a sequence of two time delayed ($\tau$) laser pulses, hereafter referred to as 'pulse-1' and 'pulse-2', a coherent superposition of the excited-state wavefunctions is generated: $\psi(t,\tau) = \chi^{(1)}(t) + \chi^{(2)}(t-\tau)$, where $\chi^{(1)}$ and $\chi^{(2)}$ are the excited-state wavefunctions generated by pulse-1 and pulse-2, respectively, as schematically depicted in Fig. 3A. The quantum interference between these two wavefunctions results into an oscillatory modulation in the population of the excited states (*9*) (Fig. 3A). The photocurrent ($I_{Photo}(\tau)$) generated by this quantum interference is proportional to the population of the excited states:

$$I_{Photo}(\tau) \propto \int_{-\infty}^{\infty} \psi(t,\tau)^* \psi(t,\tau) dt = \int_{-\infty}^{\infty} \left[ \chi^{(1)}(t)^* \chi^{(1)}(t) + \chi^{(2)}(t-\tau)^* \chi^{(2)}(t-\tau) + 2\chi^{(1)}(t)^* \chi^{(2)}(t-\tau) \right] dt, \quad (1)$$

where the first two terms are time-delay invariant, while the last term leads to an interferometric autocorrelation between the wavefunctions excited by the two laser pulses. A constructive or destructive interference of the quantum paths is obtained depending on the delay between the two pulses. This leads to oscillations in photocurrent with a frequency equal to the energy difference between the ground and excitonic states (*9*). Measuring this term in our experiments, which is imprinted onto the photocurrent, provides the information of the temporal evolution and decoherence dynamics of the wavefunction in the excited state of the molecule.

The temporal evolution of the photocurrent measured as a function of the delay between pulse-1 and pulse-2 with the nanotip positioned on the lobe of the molecule is shown in Fig. 3B. Ultrafast oscillations in the photocurrent with a period of ~ 2.5 fs persist for time-delays longer than 200 fs between the pulses, as shown in the lower inset in Fig. 3B. Figure 3C shows the comparison of the molecular spectrum obtained by STM electroluminescence measured without laser illumination with the fast Fourier transformation (FFT) of the photocurrent time trace in Fig. 3B. The spectral overlap of the peaks in the two spectra suggests that the time trace of the photocurrent is an outcome of the quantum interference measurement as discussed above and thus, it reveals the spectral profile of the molecular excitonic state formed by single-photon excitation in CuNc (Fig. 3C). To ensure that these oscillations indeed reflect the quantum excitonic interference in CuNc, we recorded the photocurrent time trace on top of bare NaCl (Fig. 3D). Here, the oscillations in the photocurrent last only for ~ 15 fs, which is roughly the duration of the incident pulses. The FFT of this time trace on NaCl reveals the shape of the spectrum of the local electric field of the pulses in the junction, which is relatively narrower compared to the far-field spectrum of the laser pulses (red curve in Fig. 3E), possibly due to the local plasmonic response.

The loss of optically induced coherence owing to the intrinsic electronic property of CuNc and its interaction with the plasmonic cavity in the STM manifests as a reduction in the amplitude of the time-resolved oscillatory photocurrent signal in Fig. 3B. A two-level density matrix calculation accounting for excitonic decoherence shows that the photocurrent would evolve as (*9*):

$$I_T(\tau) \propto \cos\left(\frac{\Delta E}{\hbar}\tau\right) \exp\left(-\frac{\tau}{T_2}\right), \quad (2)$$

where $\Delta E$ is the energy gap between the ground and the excited state, $\hbar$ is the reduced Planck's constant, and $T_2$ is the decoherence (or dephasing) time. Fitting the photocurrent signal with the expression in Eq. (2) exhibits a $T_2$ of ~ 70 fs (see Fig. S3A in supplementary materials). Notably, a faster decoherence time of only ~ 50 fs is measured for a CuNc molecule present on a thinner decoupling layer (2ML) of NaCl compared to measurements on the thicker decoupling layer of 3ML (see Fig. S2 in supplementary materials).

The electronic coherence time of CuNc is considerably shorter than the spin coherence time in single molecules (*35*). Nevertheless, a fast quantum control of population in the excited state and its readability by measuring the photocurrent in an STM opens the prospects of exploring light-wave or petahertz electronics at the single molecule level (*36*).

**Atomic-scale control of dark and bright triplet excitons**

Intermolecular interactions play an important role in coherent exciton transport, enabling electronic energy and charge transfer in photosynthesis and photovoltaic devices. Such intermolecular interactions can either be detrimental or favorable to exciton coherence and henceforth, can enhance or quench the involved processes. Here, we investigate the consequence of intermolecular interactions on exciton dynamics at the atomic length scale. By utilizing tip manipulation techniques in the STM, we assembled a dimer of CuNc molecules (inset in Fig. 4A). The time-resolved variation of the photocurrent measured as a function of the delay between pulse-1 and pulse-2 at two distinct positions of the nanotip on the dimer is shown in Fig. 4A, annotated as P1 (red cross) and P2 (blue cross) in the inset STM image. The measured time traces differ in their envelopes at longer delays (~100 fs), as indicated by the vertical gray arrows in Fig. 4A. These differences in the time traces are further underscored in the FFT spectra, as shown in Fig. 4B, along with the reference spectrum for a CuNc monomer (Fig. 3C). Unlike the single peak measured in the spectrum for the CuNc monomer, the dimer spectra exhibit multiple spectral peaks seemingly arising because of the splitting of the monomer excited states due to the intermolecular coupling.

The absorption spectrum of a CuNc molecule calculated by time-dependent density functional theory (TDDFT) shows a single excitonic peak (Fig. 4C), arising from two degenerate excited states, which have a doublet spin multiplicity (see supplementary materials for details). The intermolecular interactions in the dimer, whose electronic ground state is a spin triplet, lifts the degeneracy of the excited state, giving rise to four excited states with triplet multiplicity. Among them, only two possess non-zero optical oscillator strengths. As a consequence, the simulated light absorption spectrum of the dimer (green curve in Fig. 4C) exhibits two non-degenerate triplet excitonic peaks (E1 and E3).

Nevertheless, in the experiments four distinct spectral peaks were measured, two closely separated peaks at the nanotip position P2 and two widely separated peaks at the position P1. This spatial variation indicates site-dependent excitonic coupling within the dimer.

The electric field of the exciting laser pulses is strongly enhanced beneath the apex of the nanotip, thus, making the field-distribution highly localized and non-uniform over the molecules. This invokes the need to consider the variation of oscillator strengths and the transition matrix elements on the sub-molecular scale (see supplementary materials for details). The transition dipole, ***d***, is a vector quantity, whose components are given by:

$$d_\alpha = \left( \int_{-\infty}^{\infty} \int_{-\infty}^{\infty} \int_{-\infty}^{\infty} \Psi_i^*(x,y,z) \, \alpha \, \Psi_f(x,y,z) \, dx \, dy \, dz \right)_{\alpha=x,y,z} \quad (3)$$

where $\Psi_i$ is the initial electronic state (usually the ground state), and $\Psi_f$ is the final electronic state. The electronic response of the system to the incoming light is obtained by computing the scalar product of ***d*** and the light electric field ***E***, whereas the square modulus of ***d*** is proportional to the oscillator strength (see supplementary materials for further details and the generalization to the many-electron case). A space resolved version of the transition dipole matrix element is obtained by isolating the integrand $\mu_\alpha(x,y,z) = \Psi_i^*(x,y,z)\alpha\Psi_j(x.y,z)$. Figure 4D shows the sub-molecular scale variation of the computed transition matrix elements corresponding to the four triplet excited states.

Modelling the local electric field of the laser pulses in the STM junction using two point-dipoles (one in the nanotip and the other one in the substrate) and considering the spatially resolved differential matrix elements results into sub-molecular scale variation of the oscillator strengths. The spectra simulated for the P1 and P2 positions of the nanotip over the molecules, indicated by transparent red and blue circles in Fig. 4D, are shown in Fig. 4C (red and blue curves, respectively). The optically dark states (E2 and E4) are now apparent in the absorption spectra at the two positions. The integral of the transition matrix elements over the whole molecular volume vanishes for the dark states due to the symmetry of the molecules. However, the local excitation facilitated by the nanotip leads to symmetry breaking, thus, making the dark excitonic states bright.

The simulated dark states E2 and E4 likely correspond to the spectral peaks measured at ~ 1680 meV and ~ 1698 meV in the experiments, respectively, whereas the bright states E1 and E3 correspond to the peaks measured at ~ 1671 meV and ~ 1693 meV, respectively. It is worth mentioning that even though the simulations qualitatively agree with the spectra measured for the monomer and dimer, nonetheless, their energies show notable deviations, due to limitations in the TDDFT simulations (*37*).

The decoherence time as retrieved from the time-resolved measurements for the bright triplet excitonic state, E1, is ~ 60 fs, whereas for the dark excitonic state, E4, is ~ 50 fs (see Fig. S3B and S3C in supplementary materials), both of which are faster compared to the ~ 70 fs coherence time observed for the doublet excitons in the monomer (Fig. 3B). This is also evident from the relatively broader spectral linewidths of the FFT spectra of the dimer compared to the monomer, as shown in Fig. 4B. Estimation of the decoherence times for the states E2 and E3 is difficult owing to their closely separated spectral positions. The faster decoherence time of the triplet excitons suggests that the intermolecular interactions contribute to the quantum decoherence of the electronic states. Moreover, a strong exciton-vibronic coupling could contribute to the quenching of the coherence in CuNc molecules (*38, 39*).

The interplay between dark and bright excitons plays a crucial role in optimizing the performance of functional photovoltaics. Here, we demonstrate selective excitation of dark and bright triplet excitonic states in a CuNc dimer at the atomic scale. Figure 5A and 5B show the evolution of the FFT spectra obtained from the time-resolved photocurrent measurements at equidistant positions of the nanotip along two different axes of the dimer. The measurement positions of the nanotip are annotated as color-coded dots over the STM topographic images in the insets of Fig. 5A and Fig. 5B. Along the long-axis in the dimer (Fig. 5A), the bright excitonic state, E1, was primarily measured at the edges, whereas at the center of the dimer, the optically dark state, E4, was detected. The E1 and E4 states are energetically separated by ~ 27 meV. On the other hand, along the dotted blue line in Fig. 5B, a complex interplay of the spectral intensities of the dark, E2, and bright, E3, states was observed. E2 and E3 states are separated by ~ 13 meV.

The spatial dependence of the excitonic states indicates the significance of local field distribution and coherent molecular coupling, and highlights a pathway for atomic-scale control over excitons and their dynamics in coupled molecules.

The FFT spectra recorded at the equidistant positions of the nanotip along one axis of the monomer are shown in Fig. 5C. Here, the spectral invariance of the peak position (E0) can be distinctly seen. The measured decoherence time is also spatially invariant. Nevertheless, the spectral intensity of the peak undergoes a modulation over the molecule, its value being maximal at the edges and minimal at the center. The spatial distribution of the differential form of the transition matrix elements for a monomer (Fig. 4D) attain their maximal values only at the edges of the molecule, whereas in the center their intensities are negligible, thus, corroborating the measured variation of the spectral intensity. A higher intensity on one edge of the monomer compared to the other edge in Fig. 5C is likely due to the asymmetry of the nanotip. The spectral amplitude of the excitonic peak in this case is affected primarily by the local atomic-scale inhomogeneity of the valence electron density distribution in the molecule.

The atomic-scale variation of the intensity of the dark and bright states in the dimer is a consequence of the convolution of the local radial electric fields generated by the nanotip and the differential matrix elements (*8, 40, 41*) (Fig. 4D). The investigation of the dark states is often complicated by the need of external electric or magnetic fields to enforce their blending with the bright states, nevertheless, in the atomistic near fields of the nanotip, they can be easily accessed and their dynamics probed.

**Concluding Remarks**

The capability demonstrated in the current work of probing orbital-resolved femtosecond coherences of excitons in single molecules represents a step toward understanding and controlling the charge and energy transfer in light harvesting complexes at the molecular level. Atomic-scale switching between the dark and bright triplet excitonic states in a dimer could provide molecular level understanding of singlet exciton fission and long-range transport of the excitons. Moreover, femtosecond control of excited state dynamics is critical for realizing molecular-scale electronics (*42*) and possible exploration of quantum entanglement of electronic degrees of freedom in molecules with long-lived electronic coherences.

**Acknowledgments:** We thank Wolfgang Stiepany and Marko Memmler for technical support. FM acknowledges the Ministerio de Ciencia e Innovación projects PID2022-138288NB-C31 (MCIN/AEI/ 10.13039/501100011033/FEDER UE), the "Severo Ochoa" Programme for Centres of Excellence in R&D (CEX2020-001039-S), and the "María de Maeztu" Programme for Units of Excellence in R&D (CEX2018-000805-M). MP acknowledges partial financial support by the Centro Nazionale di Ricerca in High-Performance Computing, Big Data and Quantum Computing, PNRR 4 2 1.4, CI CN00000013, CUP H23C22000360005. All calculations were performed at the Mare Nostrum Supercomputer of the Red Española de Supercomputación (BSC-RES), the Centro de Computación Científica de la Universidad Autónoma de Madrid (CCC-UAM), and the 'Newton' and 'Alarico' HPC machines of the Physics Department at Università della Calabria.

**Author contributions**: Y.L., S.S. and M.P. contributed equally to this work. Y.L., S.S., C.C., K.K., and M.G. built the experimental setup, performed the experiments and analyzed the experimental data. M.P., and F.M. designed and performed the theoretical calculations and analyzed the theoretical data. M.G. conceived the project and designed the experiments. All authors interpreted the results and contributed to the preparation of the manuscript.

**Competing interests:** The authors declare no competing interests.

**Data and materials availability:** The data supporting the findings of this study are provided in the main text or the supplementary materials.

**Supplementary Materials**

Materials and Methods

Supplementary Text

Figs. S1 to S8

References (*1–8*)

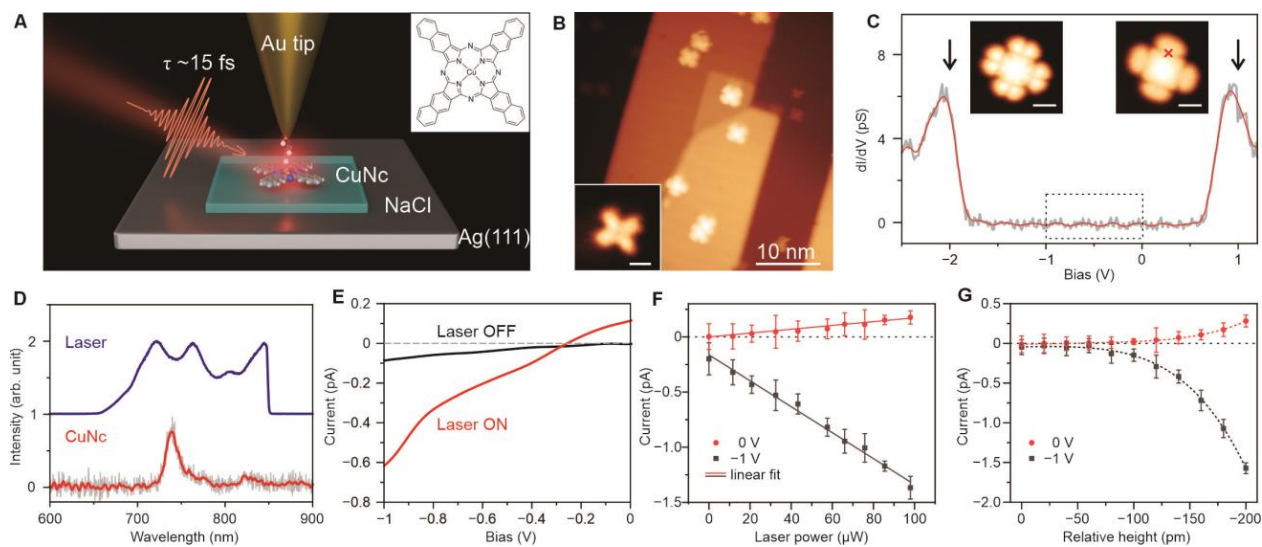

**Fig. 1. Ultrafast photon-induced electron tunneling in single molecules.** (**A**) Schematic illustration of the ultrafast photon-induced tunneling current generation in a single copper naphthalocyanine (CuNc) molecule electronically decoupled from the Ag(111) surface by an ultrathin film of sodium chloride (NaCl). Inset shows the chemical structure of the CuNc molecule. (**B**) STM image of CuNc molecules deposited on a three-monolayer-thick (3ML) film of NaCl grown on Ag(111) surface; measured at bias voltage of 1 V and constant tunneling current of 2 pA. The inset shows the topography of a single CuNc molecule (−1 V and 2 pA, scale bar: 1 nm). (**C**) Differential conductance (d$I$/d$V$) spectrum of a single CuNc molecule, measured at the nanotip position indicated by the red cross in the inset STM image. The insets show the topographic images measured at the bias of −2 V and 1 V, corresponding to the highest occupied molecular orbital (HOMO) and lowest unoccupied molecular orbital (LUMO), respectively. Scale bars: 1 nm. (**D**) Comparison of the STM-induced luminescence (STML) spectrum of a CuNc molecule (red curve) with the spectrum of the incident laser pulses (blue curve). STML spectrum was measured on the CuNc lobe (red cross in the inset of **C**) at −2.2 V and 50 pA. (**E**) Comparison of the variation of the tunneling current as a function of the applied bias (*I-V* curve) in presence (red curve) and absence (black curve) of the laser pulses in the bias range from −1 V to 0 V, as denoted by the dashed gray rectangle in **C**. (**F** and **G**) Variation of the photon-induced tunneling current as a function of increasing power of the laser pulses (**F**) and reducing nanotip height (**G**), respectively. The laser power dependence was measured for two bias voltages of 0 V (red dots) and −1 V (black dots). Solid (dashed) red and black lines denote the linear (exponential) fittings of the experimental data. Before the measurements in (**E**, **F** and **G**), the nanotip was stabilized on the molecular lobe (red cross in (**C**)) at 1 V and 2 pA, then approached to the molecule by 160 pm with open feedback loop. The laser power was set to be 50 μW in (**E** and **G**). The error bars in (**F** and **G**) are standard deviations from three consecutive measurements.

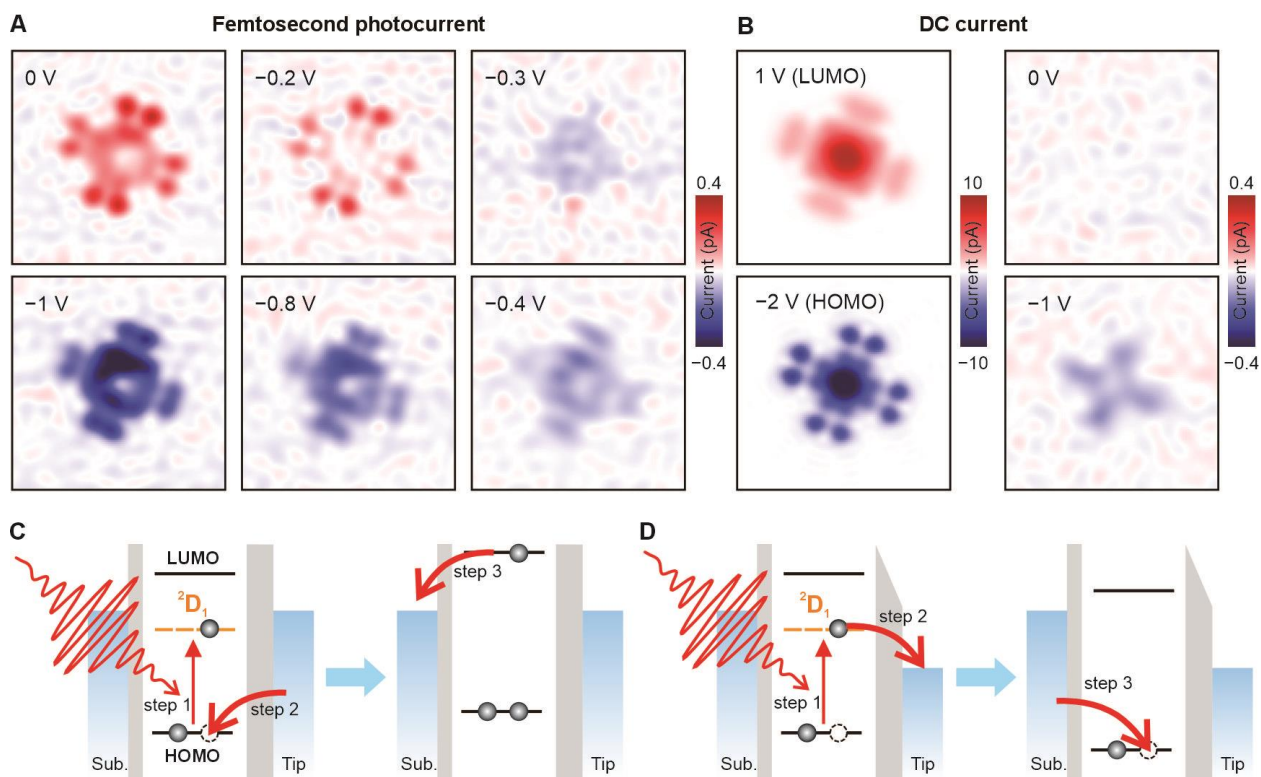

**Fig. 2: Orbital-resolved imaging of ultrafast photon-induced tunneling current.** (**A**) Ultrafast photocurrent images (4.0 nm × 4.0 nm) of a single CuNc molecule on a 3ML NaCl acquired with the open feedback loop (constant-height mode) at various biases as annotated in each individual panel. A laser power of 100 μW was used in the measurements. Before acquiring the images, the nanotip was stabilized at 1 V and 2 pA on the molecular lobe (red cross in Fig. 1C), and then approached by 140 pm. (**B**) DC tunneling current images (4.0 nm × 4.0 nm) of the CuNc molecule in absence of laser pulses acquired at various biases in the constant height mode of the STM. The DC current images at 1 V and -2 V were measured after stabilizing the nanotip on the molecular lobe (red cross in Fig. 1C) at 1 V and 2 pA. For the DC current images at 0 V and -1 V, the nanotip was stabilized at 1 V and 2 pA on the molecular lobe and further approached by 140 pm before the measurements to enable a direct comparison with the images in **A**. (**C** and **D**) Schematic description of the mechanism of ultrafast photocurrent generation from a single CuNc molecule at 0 V and −1 V, respectively. The red arrows indicate the electron transfer process via the molecular orbitals.

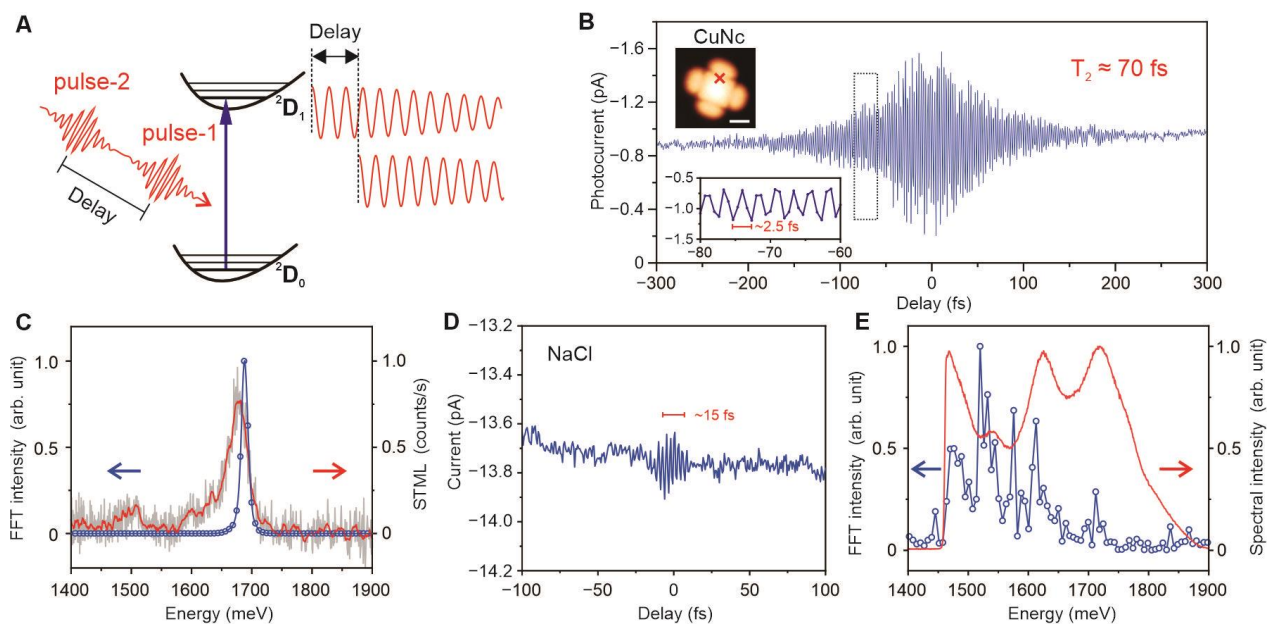

**Fig. 3: Quantum decoherence of the excitons in a single molecule.** (**A**) Schematic illustration of the excitonic wavepacket interferometry. Excitonic wavepackets launched by two time-delayed laser pulses ('pulse-1' and 'pulse-2') interfere in the excited electronic state. (**B**) Ultrafast photon-induced tunneling current measured on the molecular lobe (red cross in the inset) as a function of the delay between pulse-1 and pulse-2. The inset indicates the modulation of the tunneling current in the small-time window between −80 to −60 fs. The nanotip was stabilized at 1 V and 2 pA on the molecular lobe and was further approached by 200 pm before starting the time-resolved measurements. The power of each laser pulse was set to be 50 μW and the acquisition time per data point was 50 ms. (**C**) Comparison of the molecular spectrum obtained with STML (red curve) and by fast Fourier transformation of the time trace in **B** (blue curve). STML spectrum was measured on the CuNc lobe at −2.5 V and 50 pA. (**D**) Time trace of ultrafast photocurrent measured on bare NaCl; width of the cross-correlation curve reveals the time resolution in the experiments, which is approximately 15 fs. (**E**) Comparison of the spectrum of the incident laser pulses (red curve) with the spectrum obtained by fast Fourier transformation of the time trace in **D** (blue curve) on NaCl.

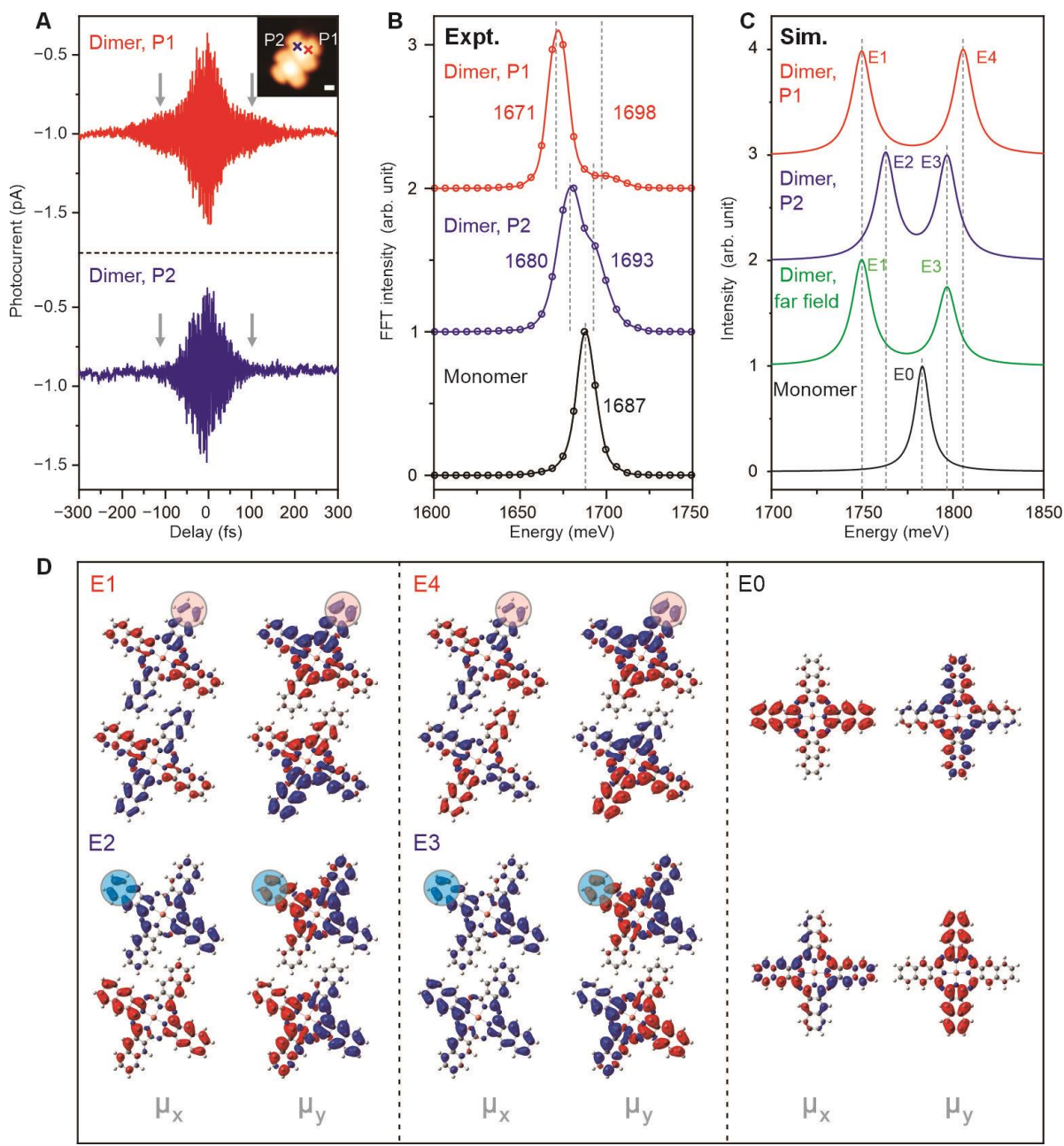

**Fig. 4: Femtosecond coherence of triplet excitons in coherently coupled molecules. (A)** Time-resolved ultrafast photocurrent traces measured at the positions marked by red ('P1') and blue ('P2') crosses in the inset STM image of a CuNc dimer measured at 1 V with a constant tunneling current of 2 pA. Scale bar: 1 nm. Vertical gray arrows indicate the differences in the two time-traces at delays of ~ 100 fs. The photocurrent measurements were conducted after stabilizing the nanotip on the molecular lobe at 1 V and 2 pA and then further approached by 200 pm. The power of each laser pulse was set to be 50 μW. **(B)** Comparison of the spectra obtained by FFT of the measurement in **A** with the spectrum of a single CuNc molecule (Fig. 3C). Dashed gray lines indicate the shifting and splitting of the spectral peaks in the dimer with respect to the monomer. **(C)** Comparison of the spectra retrieved from TDDFT simulations for a monomer (black curve) with the spectra of the dimer in the far-field (green curve) and near-field simulated at the nanotip positions of P1 (semi-transparent red circle in panel **D**) and P2 (semi-transparent blue circle in **D**)
Page **16** of 17

over the dimer (identical to the red and blue crosses in the inset in **A**). The near-field spectra were simulated by considering a radial distribution of electric field under the apex of the nanotip of the STM generated by two-point dipoles, one along the nanotip axis and the other on the Ag(111) substrate. **(D)** Spatial variation of the differential form of the transition matrix elements for the four relevant optical transitions (E1, E2, E3 and E4) in the dimer (left and middle panels) and the optical transition (E0) in the monomer (right panel) for dipoles pointing in the *x* and *y* directions. Only the top-views of the $\mu_x$ and $\mu_y$ isosurfaces (red 0.001 au, blue -0.001 au, au: atomic units) are given. $\mu_x$ and $\mu_y$ vanish at the molecular plane and are even for z→-z transformations. In all cases $\mu_z$, which is odd for z→-z, is much smaller (by at least one order of magnitude) with respect to $\mu_x$ and $\mu_y$ (see the supplementary materials for details). Semi-transparent red and blue circles indicate the position of the nanotip used for simulating the corresponding color-coded spectra in **C**.

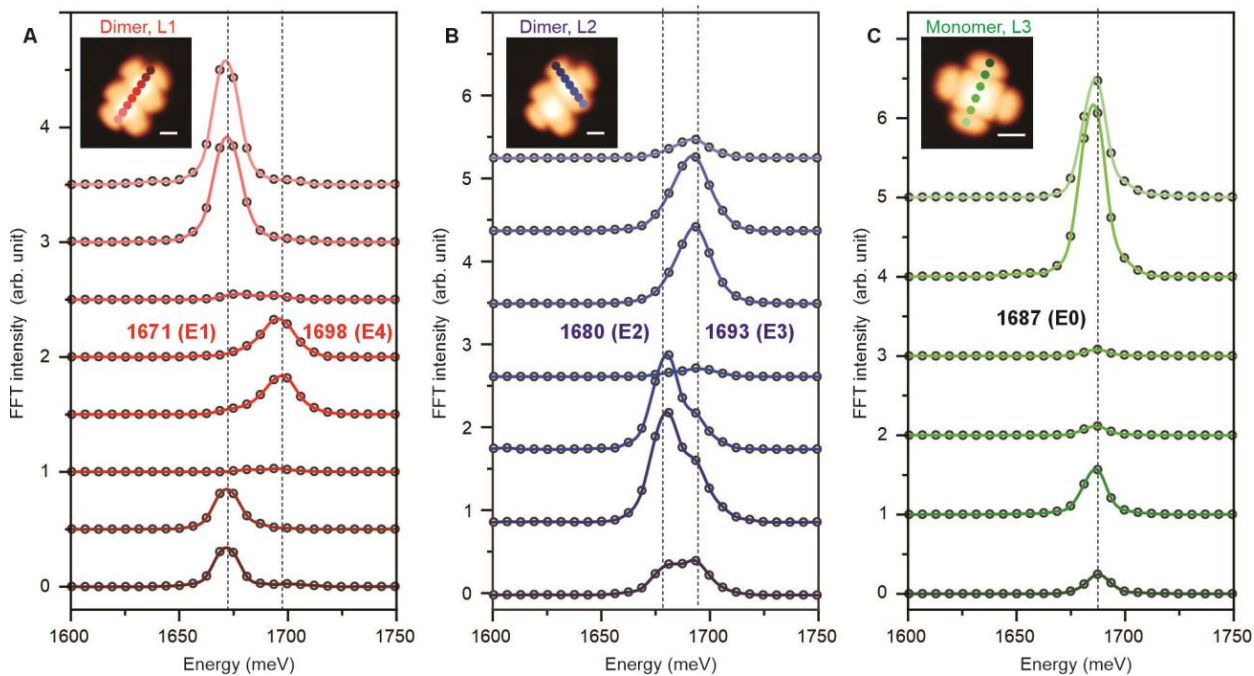

**Fig. 5: Selective excitation of dark and bright triplet excitons at the atomic length scale. (A, B)** A series of FFT spectra retrieved from time-resolved photocurrent measurements at eight equidistant positions of the nanotip over the CuNc dimer. Insets show the STM images of a CuNc dimer measured at 1 V with a constant tunneling current of 2 pA. Scale bar: 1 nm. The color of the spectra denotes the position of the nanotip over the dimer as annotated by the corresponding color-coded dots in the STM images in the inset. Atomic-scale movement of the nanotip over the dimer leads to switching of the spectral intensity between dark (E2 and E4) and bright (E1 and E3) triplet excitonic states. **(C)** Same as **A** and **B** for the measurements on a monomer. The spectral position of the doublet exciton is insensitive to the position of the nanotip over the molecule, only its intensity undergoes spatial modulation. The spectra have been vertically shifted for clarity.

Supplementary Materials for

# Orbital-resolved imaging of coherent femtosecond exciton dynamics in coupled molecules

Yang Luo[1,2]†, Shaoxiang Sheng[1,3]†, Michele Pisarra[4,5]†, Caiyun Chen[1], Fernando Martin[6,7]*, Klaus Kern[1,8], Manish Garg[1],*

[1]Max Planck Institute for Solid State Research, Heisenbergstr. 1, 70569 Stuttgart, Germany

[2]Hefei National Laboratory, University of Science and Technology of China, Hefei, 230088, China

[3]Tsientang Institute for Advanced Study, Hangzhou, 310024, China

[4]Dipartimento di Fisica, Università della Calabria, Via P. Bucci, cubo 30C, 87036, Rende (CS), Italy

[5]INFN-LNF, Gruppo Collegato di Cosenza, Via P. Bucci, cubo 31C, 87036, Rende (CS), Italy

[6]Instituto Madrileño de Estudios Avanzados en Nanociencia (IMDEA Nano), Faraday 9, Cantoblanco, 28049 Madrid, Spain

[7]Departamento de Química, Módulo 13, Universidad Autónoma de Madrid, 28049 Madrid, Spain

[8]Institut de Physique, Ecole Polytechnique Fédérale de Lausanne, 1015 Lausanne, Switzerland

*Corresponding authors. Email: mgarg@fkf.mpg.de and fernando.martin@imdea.org

†These authors contributed equally to this work: Y. Luo, S. Sheng and M. Pisarra

**The PDF file includes:**

    Materials and Methods
    Supplementary Text
    Figs. S1 to S8
    Tables S1 and S2
    References

## Materials and Methods

Sample and tip preparation

The experiments were conducted in a home-built scanning tunneling microscope (STM) operating in ultra-high vacuum conditions (~ $5\times10^{-11}$ mbar) and at liquid Helium temperature (~ 11 K). The Ag(111) substrate was prepared by repeated cycles of sputtering with Ar+ ions (~1.0 keV) and thermal annealing (~ 400 °C). Au tips prepared by electrochemical etching were used in all the experiments. Copper napthalocyanine (CuNc) molecules were thermally evaporated onto the NaCl covered Ag(111) surface using a homemade evaporator while maintaining the substrate at ~ 11 K.

Optical Set-up

An ultra-broadband Ti:Sa oscillator with a repetition rate of ~ 80 MHz and a spectral range spanning from 650 to 1100 nm was used in the experiments. The laser pulses were focused onto the STM junction by a biconvex lens of 7.5 cm focal length, which was mounted inside the UHV chamber. The spectral chirp introduced by the window of the STM, biconvex lens and the propagation of laser pulses in air was precompensated by multiple reflections off the surface of a pair of chirped dielectric mirrors, with a group-delay-dispersion (GDD) of ~ −60 $fs^2$. A second harmonic generation based fringe resolved autocorrelator (FRAC) with ~ 20 μm thick BBO crystal was used to measure the duration of the laser pulses with identical dispersion as in the optical path to the STM junction. The laser pulses were spectrally filtered in the range of ~ 650-850 nm by a shortpass filter before being directed to the STM junction.

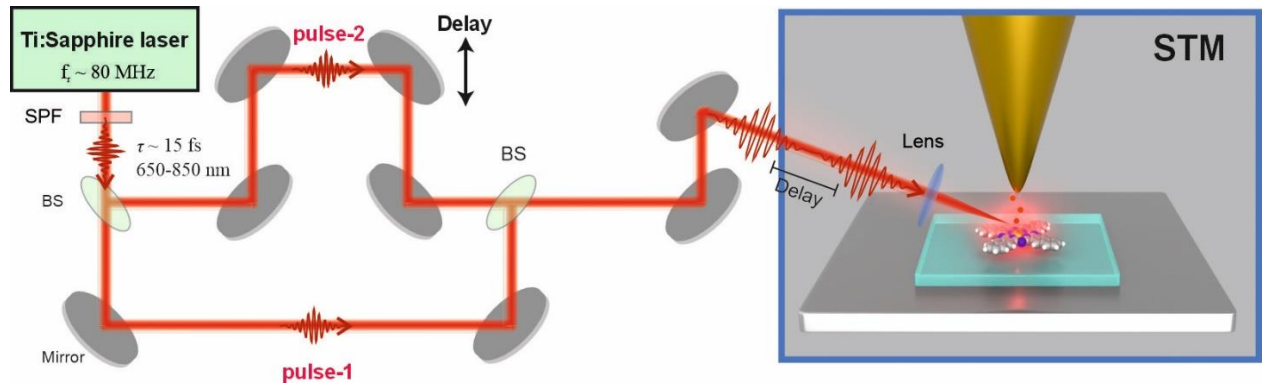

**Fig. S1. Experimental setup.** In the time-resolved measurements, two time-delayed laser pulses (pulse-1 and pulse-2) in the spectral range spanning from ~ 650 nm to 850 nm were generated by traversing the ultra-broadband laser pulses from a Ti:Sa oscillator through an ~ 850 nm shortpass filter. The laser pulses were combined and focused onto the STM junction by a biconvex lens of focal length of 7.5 cm. BS: Beam splitter; SPF: Shortpass filter.

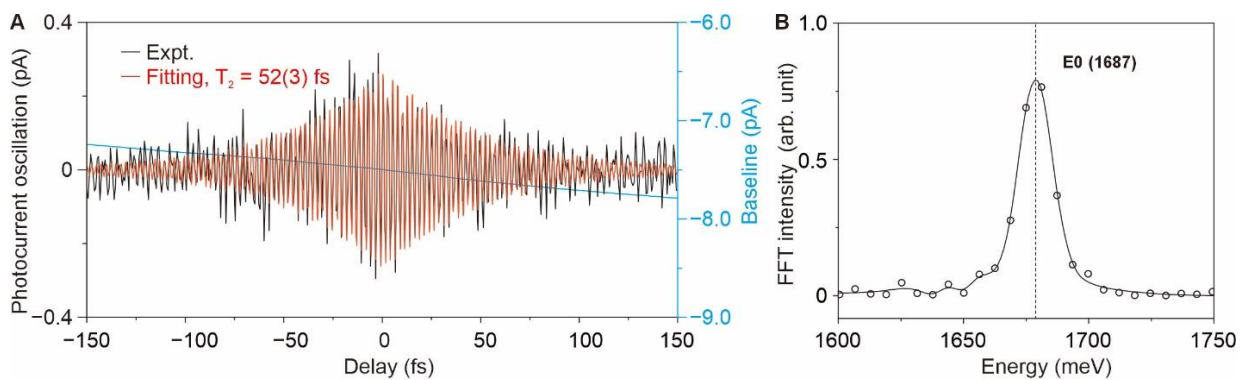

**Fig. S2. Quantum decoherence of excitons in a single CuNc molecule on a 2 ML NaCl on the Ag(111).** (**A**) Ultrafast photon-induced tunneling current measured on the lobe of a single CuNc molecule present on top of a 2ML thick NaCl island as a function of the delay between pulse-1 and pulse-2. The oscillatory component (black curve) of the photocurrent was extracted by subtracting a smoothed baseline (blue curve) from the raw time trace. The nanotip was stabilized at 1 V and 2 pA on the molecular lobe and was further approached by 260 pm before starting the time-resolved measurements. The decoherence time $T_2$ was determined by fitting the time trace to the function, $I = \cos(\omega_0 \tau) * \exp(-|\tau|/T_2)$, where $\omega_0$ corresponds to the excitonic absorption peak, and $\tau$ is the time delay between the pulses. (**B**) Fast Fourier transformation (FFT) spectrum of the time trace in **A**, indicating the position of the excitonic peak in the monomer.

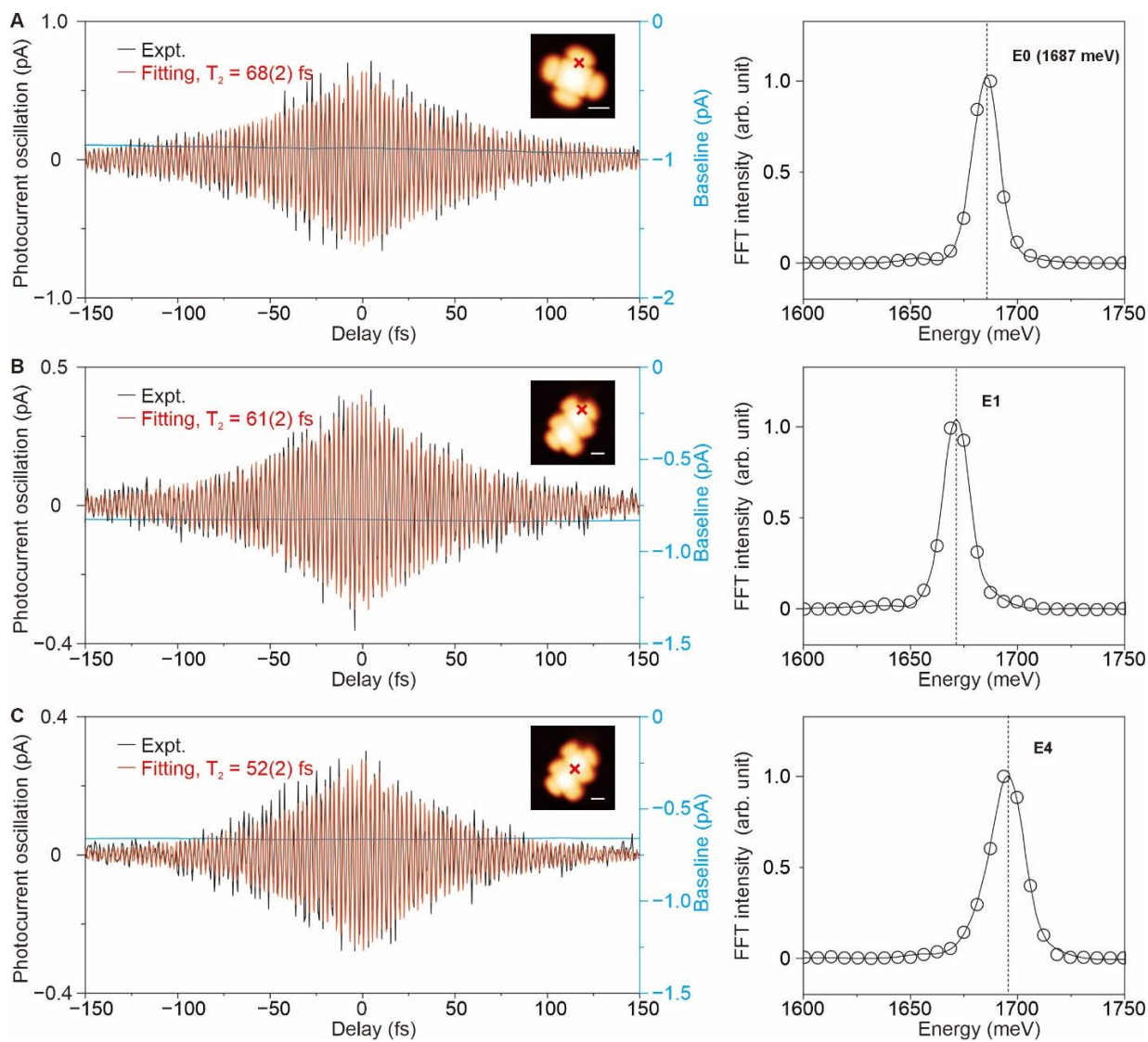

**Fig. S3. Retrieving decoherence time of excitons in monomer and dimer of CuNc molecules from the time-resolved measurements.** (**A**-**C**), Left Panels: Exciton decoherence times ($T_2$) determined by fitting the time-resolved photocurrent traces obtained with the nanotip of the STM positioned on the lobe of the monomer (A), one edge of the dimer (B), and on the center of the dimer (C). The oscillatory component (black curve) of the photocurrent time traces were extracted by subtracting a smoothed baseline (blue curve) from the raw time traces. The decoherence time $T_2$ was determined by fitting the photocurrent time trace to the function, $I = \cos(\omega_0 \tau) * \exp(-|\tau|/T_2)$, where $\omega_0$ corresponds to the excitonic absorption peak, and $\tau$ is the time delay between the pulses (pulse-1 and pulse-2). The decoherence time of a monomer is estimated to be ~ 68 fs. The lowest energy peak of molecular dimer, i.e. E1, exhibits a shorter decoherence time of ~ 61 fs, whereas the decoherence time of the highest energy peak, i.e. E4, is even shorter (~ 52 fs). The right panels in **A**, **B** and **C** display the fast Fourier transform (FFT) spectra of the corresponding time traces.

Density Functional Theory (DFT) and Time-Dependent Density Functional Theory (TDDFT) simulations

Atomistic calculations have been performed with the Gaussian 16 program package [1]. We carried out unrestricted all-electron calculations using the B3LYP functional [2] and the 6-31G(d,p) basis function, including the Grimme D3 [3] van der Waals correction (both in the dimer and in the monomer calculations for consistency). The neutral isolated Copper Naphthalocyanine (CuNc) molecule in its ground state has an unpaired electron, hence for the dimer CuNc calculation we investigated both the singlet and the triplet spin multiplicities; the triplet spin multiplicity turned out to be slightly favored in energy and was the one chosen for further analyses. The electronic excitation spectrum and transition dipole moments were calculated within a TDDFT approach [4,5] at the 6-31G-B3LYP level, obtaining up to the 30th excited state for the monomer and the 50th for the dimer. The molecular structures and graphical representations of the molecular orbitals and space dependent dipole moments were prepared using GaussView 6.1.1 [6].

CuNc monomer DFT and TDDFT results

In Fig. S4, we report the frontier orbitals (HOMO and LUMO) for the CuNc monomer. Since the electronic multiplicity of the molecule is a doublet, we have different energies for the spin up and spin down molecular orbitals (MOs). Their spatial wave function is identical in practice. Interestingly, there are two couples of degenerate LUMOs for both spin orientations. We observe that the calculated HOMO-LUMO gap is 1.872 eV for the spin up orbitals and 1.907 eV for the spin down orbitals, in excellent agreement with previously reported calculations [7].

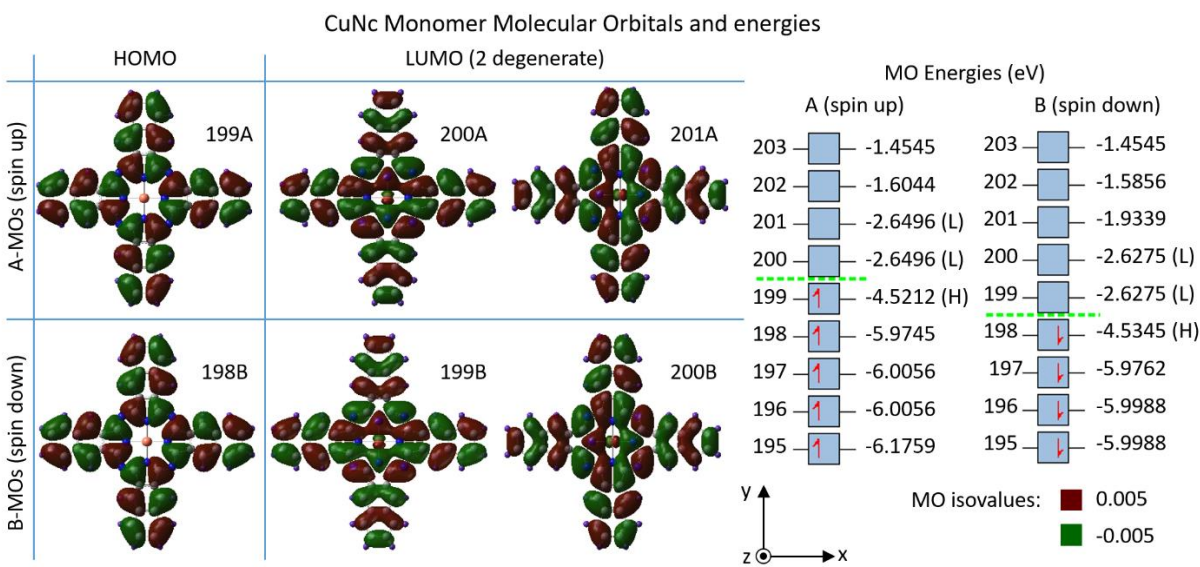

**Fig. S4**. **Frontier MOs and orbital energies of the CuNc monomer.** Only the top view is given. All the shown orbitals are odd upon reflection over the molecular plane (z=0). The orbitals are ordered in energy and assigned a cardinal number from lowest to highest and a letter, for which we follow the convention A=spin up and B=spin down. In the MO energy diagram, for each spin orientation, the HOMO and the two degenerate LUMOs are singled out by an "H" and "L", respectively. In the bottom right corner, we give the Cartesian axes orientation used in the MO picture and the color code.

| TableS1: CuNc monomer excited states as obtained by TD-DFT ||||||||
| State | Spin mult. | Energy (eV) | Energy (nm) | Transition dipole moment **d**(au) ||| Oscillator strength |
| | | | | $d_x$ | $d_y$ | $d_z$ | |
|---|---|---|---|---|---|---|---|
| 1 | 3.465 | 0.9277 | 1336 | 0 | -0.0797 | 0 | 0.0001 |
| 2 | 3.465 | 0.9277 | 1336 | -0.0803 | 0 | 0 | 0.0001 |
| **3** | **2.003** | **1.7831** | **695** | **0** | **-3.6281** | **0** | **0.5750 (E0)** |
| **4** | **2.003** | **1.783** | **695** | **-3.6275** | **0** | **0** | **0.5748 (E0)** |
| 5 | 2.006 | 1.8478 | 671 | 0 | 0 | 0 | 0 |
| 6 | 3.465 | 2.0475 | 606 | 0 | 0 | 0 | 0 |
| 7 | 3.465 | 2.1694 | 572 | 0 | 0 | 0 | 0 |
| 8 | 3.465 | 2.2775 | 544 | 0 | -0.0555 | 0 | 0.0002 |
| 9 | 3.465 | 2.2775 | 544 | -0.0557 | 0 | 0 | 0.0002 |
| 10 | 2.028 | 2.4886 | 498 | 0 | 0 | 0 | 0 |

In Table S1 we report the first 10 excited states of the CuNc monomer as obtained in a TDDFT calculation. In all cases we report the total spin multiplicity of the excited state and the energy (in eV and the corresponding conversion to nm). We also report the transition dipole moment, in atomic units (au), and the oscillator strength of the ground to excited state electronic transitions. We observe that, since the spin multiplicity of the ground state is 2, only the excited states with spin multiplicity 2 are reached upon the absorption of a photon. In the list given in Table S1 there are three such states (excited states 3, 4, and 5) that meet this criterion and have an energy comparable with the experimentally measured electronic transition. A lot of information is obtained by looking at the main MO transitions that give rise to each excited state. We found:

$$\Psi_{ES3} = 0.68854(|199A\rangle \rightarrow |201A\rangle) + 0.71299(|198B\rangle \rightarrow |200B\rangle), \quad (1)$$

$$\Psi_{ES4} = 0.68854(|199A\rangle \rightarrow |200A\rangle) + 0.71299(|198B\rangle \rightarrow |199B\rangle), \quad (2)$$

$$\Psi_{ES5} = |198B\rangle \rightarrow |201B\rangle. \quad (3)$$

Using as an example eq.1, the notation indicates that there are two main contributions to excited state 3, which is then written as a linear combination of two Slater determinants: the first (first term in eq. 1) is obtained from the ground state removing the electron from the 199A spin-orbital and putting it in the 201A spin-orbital (with an expansion coefficient 0.689); the second Slater determinant is obtained removing the electron from the 198B spin-orbital and putting it in the 200B spin-orbital (with a coefficient 0.713). The excited states 3 and 4 are a combination of HOMO-LUMO transitions. Given the symmetries of the HOMO and LUMO orbitals, these excited states have nonzero transition dipole moment for the in-plane coordinates (either $d_x$ or $d_y$), while the out-of-plane transition dipole moment $d_z$ vanishes [8]. Interestingly, the excited state 5 is a HOMO-LUMO+1 transition, which is dark and has a zero transition dipole moment for symmetry considerations. In our analysis we then selected the excited states 3 and 4 of the monomer, which are identified by the E0 label in the main text.

CuNc monomer DFT and TDDFT results

Upon the formation of a dimer of CuNc molecules the intermolecular interaction causes the formation of closely spaced frontier orbitals, derived from the parent HOMO and LUMO of the isolated CuNc molecule. This fact is made manifest in Figs. S5 and S6, where we report the frontier orbitals of a CuNc dimer in a triplet electronic configuration for the spin up (Fig. S5) and spin down (Fig. S6). For both spin orientations we find a "HOMO group" of two almost degenerate MOs, and a "LUMO group" made of four almost degenerate MOs. Also in this case, while the orbital energies differ between spin up and spin down, the spatial shape of the MO is almost indistinguishable between the corresponding orbitals.

As for the excited states of the dimer, we report in Table S2 the results for the lowest 20 excited states. Since the ground state is a triplet, only those states that have triplet spin multiplicity are accessible through light absorption. We further point out that there are six excited states (states 13 to 18) with triplet multiplicity in the 1.5 – 2.0 eV energy range. The expansion of the excited states 13 to 18 is given in eqs.4-9.

$$\Psi_{ES13} = 0.45177(|397A\rangle \to |399A\rangle) + 0.18226(|397A\rangle \to |401A\rangle) + \\ 0.47321(|398A\rangle \to |400A\rangle) + 0.11474(|398A\rangle \to |402A\rangle) + \\ 0.46890(|395B\rangle \to |397B\rangle) + 0.18886(|395B\rangle \to |399B\rangle) + \\ 0.49116(|396B\rangle \to |398B\rangle) + 0.11782(|396B\rangle \to |400B\rangle), \quad (4)$$

$$\Psi_{ES14} = -0.11278(|397A\rangle \to |400A\rangle) + 0.47827(|397A\rangle \to |402A\rangle) + \\ -0.17398(|398A\rangle \to |399A\rangle) + 0.45002(|398A\rangle \to |401A\rangle) + \\ -0.11803(|395B\rangle \to |398B\rangle) + 0.49981(|395B\rangle \to |400B\rangle) + \\ -0.17738(|396B\rangle \to |397B\rangle) + 0.46320(|396B\rangle \to |399B\rangle), \quad (5)$$

$$\Psi_{ES15} = -0.18034(|397A\rangle \to |399A\rangle) + 0.45026(|397A\rangle \to |401A\rangle) + \\ -0.11599(|398A\rangle \to |400A\rangle) + 0.47419(|398A\rangle \to |402A\rangle) + \\ -0.18542(|395B\rangle \to |397B\rangle) + 0.46554(|395B\rangle \to |399B\rangle) + \\ -0.11976(|396B\rangle \to |398B\rangle) + 0.49164(|396B\rangle \to |400B\rangle), \quad (6)$$

$$\Psi_{ES16} = 0.47412(|397A\rangle \to |400A\rangle) + 0.11103(|397A\rangle \to |402A\rangle) + \\ 0.45286(|398A\rangle \to |399A\rangle) + 0.17610(|398A\rangle \to |401A\rangle) + \\ 0.49068(|395B\rangle \to |398B\rangle) + 0.11477(|395B\rangle \to |400B\rangle) + \\ 0.46872(|396B\rangle \to |397B\rangle) + 0.18084(|396B\rangle \to |399B\rangle), \quad (7)$$

$$\Psi_{ES17} = -0.49854(|395B\rangle \to |401B\rangle) - 0.49420(|395B\rangle \to |402B\rangle) + \\ 0.50845(|396B\rangle \to |401B\rangle) + 0.49028(|396B\rangle \to |402B\rangle), \quad (8)$$

$$\Psi_{ES18} = -0.49721(|395B\rangle \to |401B\rangle) + 0.50136(|395B\rangle \to |402B\rangle) + \\ -0.48752(|396B\rangle \to |401B\rangle) + 0.50538(|396B\rangle \to |402B\rangle). \quad (9)$$

If we follow the convention to consider as HOMO the MOs belonging to the HOMO group of the dimer and as LUMO the MOs belonging to the LUMO group, then a quick inspection of ES17 and ES18 demonstrates that they are formed by HOMO-LUMO transitions. For this reason, we focus on ES13, ES14, ES15, and ES16, which are all obtained as linear combinations of HOMO-LUMO transitions. In Table S2, they are identified as E1, E2, E3, and E4, as in the main text. We observe that these four excited states are derived from the two E0 excited states of the monomer. The degeneracy is lifted by the intermolecular interaction so that they are split into 4 distinct states. Interestingly, while all of them are characterized by $d_z=0$ (a property derived from the "parent" E0 states), only E1 and E3 have nonzero in-plane transition dipole moments. As consequence, E1 and E3 can be populated via light absorption, whereas E2 and E4 are "dark" states (they have zero oscillator strength).

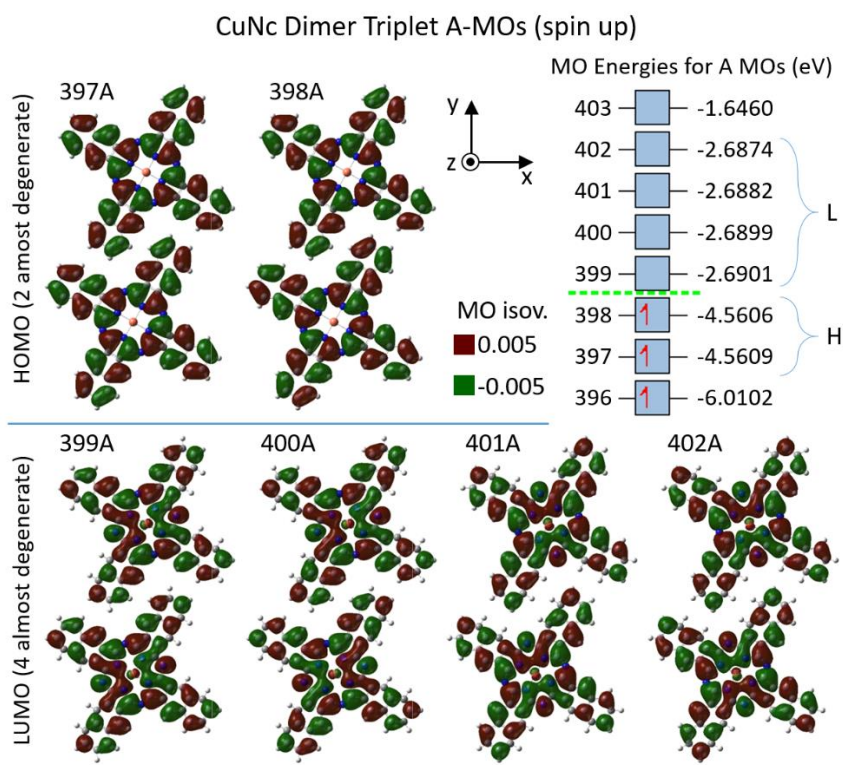

**Fig. S5**. **Frontier MOs and orbital energies of the CuNc dimer for the spin up MOs**. All the shown orbitals are odd upon reflection over the molecular plane (z=0). Only the top view is given. In the MO energy diagram, the HOMO group (H) and the LUMO group (L) are highlighted.

Page **8** of **13**

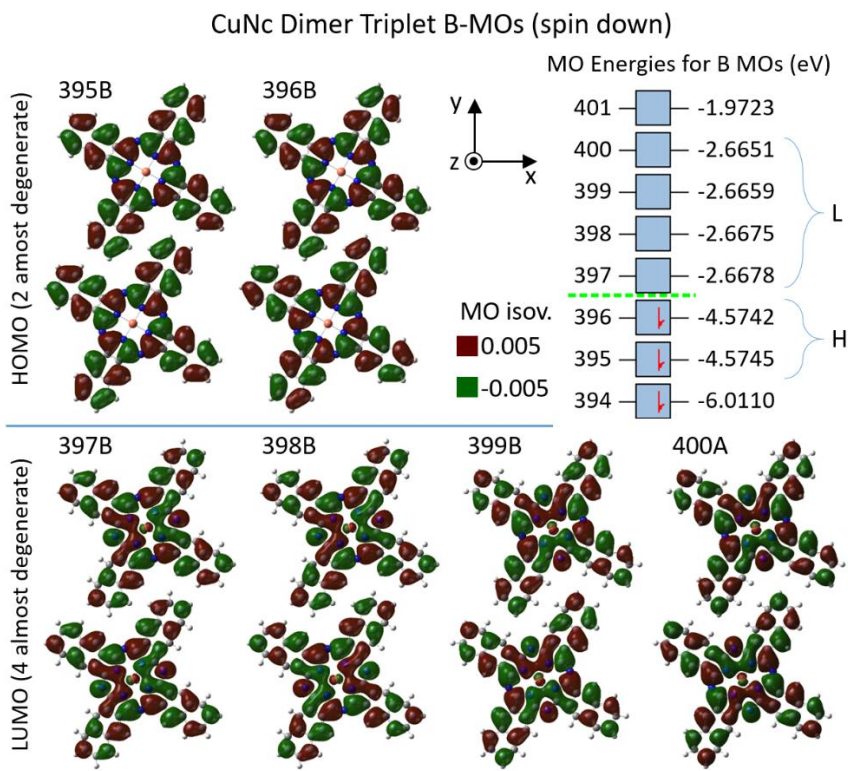

**Fig. S6. Frontier MOs and orbital energies of the CuNc dimer for the spin down MOs.** Same as for Fig.S5.

| Table2: CuNc dimer excited state as obtained by TD-DFT ||||||||
| State | Spin mult. | Energy (eV) | Energy (nm) | Transition dipole moment (au) ||| Oscillator strength |
| | | | | $d_x$ | $d_y$ | $d_z$ | |
| --- | --- | --- | --- | --- | --- | --- | --- |
| 1 | 4.125 | 0.9262 | 1339 | -0.0261 | 0.1259 | 0 | 0.0004 |
| 2 | 4.125 | 0.9262 | 1339 | -0.0043 | 0.0209 | 0 | 0 |
| 3 | 4.125 | 0.9296 | 1334 | 0.0242 | 0.008 | 0 | 0.003 |
| 4 | 4.125 | 0.9296 | 1334 | 0.1003 | 0.0332 | 0 | 0 |
| 5 | 3.608 | 1.6812 | 738 | -0.0013 | 0.0016 | 0 | 0 |
| 6 | 3.608 | 1.6812 | 738 | -0.0008 | 0.001 | 0 | 0 |
| 7 | 3.608 | 1.6869 | 735 | 0.0056 | -0.0042 | 0 | 0 |
| 8 | 3.608 | 1.6869 | 735 | -0.0084 | 0.0063 | 0 | 0 |
| 9 | 3.608 | 1.7171 | 722 | -0.0036 | 0.0036 | 0 | 0 |
| 10 | 3.608 | 1.7171 | 722 | 0.0024 | -0.0024 | 0 | 0 |
| 11 | 3.609 | 1.7228 | 720 | 0.0054 | -0.0054 | 0 | 0 |
| 12 | 3.608 | 1.7228 | 720 | -0.0145 | 0.0145 | 0 | 0 |
| **13** | **3.004** | **1.7497** | **709** | **0.2234** | **5.7351** | **0** | **1.4121 (E1)** |
| **14** | **3.004** | **1.7629** | **703** | **0.0002** | **-0.0002** | **0** | **0.0 (E2)** |
| **15** | **3.004** | **1.7967** | **690** | **4.8584** | **-0.2200** | **0** | **1.0411 (E3)** |
| **16** | **3.004** | **1.8056** | **687** | **-0.0011** | **0.0003** | **0** | **0.0 (E4)** |
| 17 | 3.006 | 1.8489 | 671 | 0 | 0 | 0 | 0 |

| 18 | 3.006 | 1.8489 | 671 | 0 | 0 | 0 | 0 |
| 19 | 4.125 | 2.0464 | 606 | 0.0011 | -0.0041 | 0 | 0 |
| 20 | 4.125 | 2.0464 | 606 | 0.0003 | -0.0011 | 0 | 0 |

Calculation of the differential form of the transition dipole matrix elements

The transition dipole matrix elements (in atomic units, which are adopted throughout unless otherwise stated) between the initial state $\Psi_i$ and the final state $\Psi_j$, whose energies are $\varepsilon_i$ and $\varepsilon_j$ are given by:

$$\boldsymbol{d}^{ij} = \sum_{q=1}^{N} \int_{\Omega} \Psi_i^*(\boldsymbol{r}_1, \boldsymbol{r}_1..\boldsymbol{r}_N) \boldsymbol{r}_q \Psi_j(\boldsymbol{r}_1, \boldsymbol{r}_1..\boldsymbol{r}_N) d^3 r_1 d^3 r_2 \ldots d^3_N, \tag{10}$$

where, $N$ is the number of electrons in the system and $\Omega$ is the total volume. We will assume that $\Psi_i$ is the ground state, which is written as a Slater determinant of the $N$ occupied spin orbitals. The transition dipole matrix elements $\boldsymbol{d}^{ij}$, are connected to the oscillator strength of the excited state through the relation, $f^{ij} = \frac{2}{3}(\varepsilon_j - \varepsilon_i)|\boldsymbol{d}^{ij}|^2$. In what follows, we will assume that the excited state $\Psi_j$ can be written as an expansion of $M$ single electron excitations in which an electron is transferrered from one spin orbital in the ground state $|g\rangle$ to a spin orbital in the excited state $|e\rangle$, (eqs. 1-9), i. e.

$$\Psi_j = \sum_{l=1}^{M} c_l^j (|g_l\rangle \to |e_l\rangle), \tag{11}$$

In this way, the Cartesian components of the transition dipole moment take the form:

$$d_x^j = \sum_{l=1}^{M} c_l^j \int_{-\infty}^{\infty} \int_{-\infty}^{\infty} \int_{-\infty}^{\infty} \varphi_{g_l}^*(x,y,z) \, x \, \varphi_{e_l}(x,y,z) dx \, dy \, dz. \tag{12}$$

$$d_y^j = \sum_{l=1}^{M} c_l^j \int_{-\infty}^{\infty} \int_{-\infty}^{\infty} \int_{-\infty}^{\infty} \varphi_{g_l}^*(x,y,z) \, y \, \varphi_{e_l}(x,y,z) dx \, dy \, dz. \tag{13}$$

$$d_z^j = \sum_{l=1}^{M} c_l^j \int_{-\infty}^{\infty} \int_{-\infty}^{\infty} \int_{-\infty}^{\infty} \varphi_{g_l}^*(x,y,z) \, z \, \varphi_{e_l}(x,y,z) dx \, dy \, dz. \tag{14}$$

where $\varphi$ denotes a spin orbital. In eqs. 12-14 we dropped the $i$ label and took advantage of the orto-normality property of the spin orbitals, to carry out the integration in eq.10. Ultimately, the dipole transition matrix elements are written as a "simple" sum of standard one body integrals. As observed in the main text, it is possible to extract a differential form of the transition dipole matrix elements, by isolating the integrand functions in eqs. 12-14. Given the final excited state, labelled with the letter $j$, we are left with three 3D functions, i. e.:

$$\mu_x^j(x,y,z) = \sum_{l=1}^{M} c_l^j \, \varphi_{g_l}^*(x,y,z) \, x \, \varphi_{e_l}(x,y,z), \tag{15}$$

$$\mu_y^j(x,y,z) = \sum_{l=1}^{M} c_l^j \, \varphi_{g_l}^*(x,y,z) \, y \, \varphi_{e_l}(x,y,z), \tag{16}$$

$$\mu_z^j(x,y,z) = \sum_{l=1}^{M} c_l^j \, \varphi_{g_l}^*(x,y,z) \, z \, \varphi_{e_l}(x,y,z), \tag{17}$$

which are easily evaluated, known the expansion coefficients $c_i^f$, and the spatial form of the spin orbitals $\varphi$. In our cases we obtained the spin orbitals (and hence the $\boldsymbol{\mu}^f$) on a spatial grid (spacings $\frac{1}{6}$ au) using the cubegen utility within the Gaussian package. In Fig. S7 and Fig. S8 we report the $\mu_z$ for the E0 states in the monomer (Fig. S7) and the E1-E4 states in the dimer (Fig.S8).

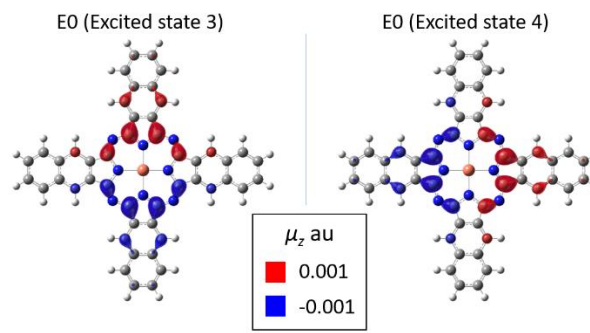

**Fig. S7** Spatial variation of the differential form of the dipole transition matrix elements $\mu_z$ (the $\mu_x$, and $\mu_y$ terms are given in Fig.4D of the main text) for the 3$^{rd}$ and 4$^{th}$ excited states (E0) of the CuNc molecule. Only the top-view is given. In both cases, $\mu_z$ is even for reflection on the molecular plane ($z \to -z$).

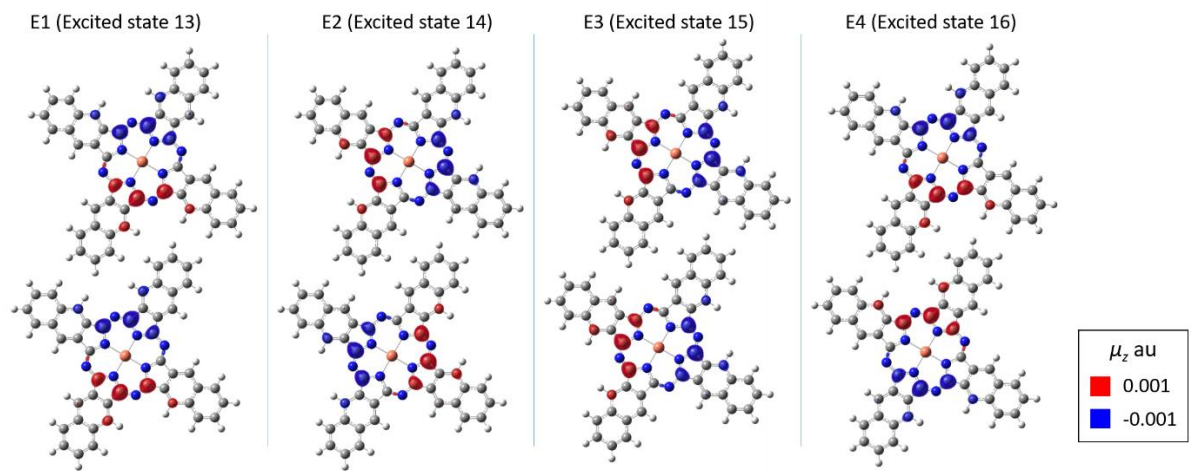

**Fig. S8**. Spatial variation of the differential form of the dipole transition matrix elements $\mu_z$ (the $\mu_x$, and $\mu_y$ terms are given in Fig.4D of the main text) for the 13$^{th}$ (E1), 14$^{th}$ (E2), 15$^{th}$ (E3), and 16$^{th}$ (E4) excited states of the CuNc dimer. Only the top-view is given. Also in these cases, $\mu_z$ is even for reflection on the molecular plane ($z \to -z$).

Page **11** of **13**

Beside a qualitative visualization of the molecular regions that are optically more active for a given excited state $j$, the expansion in eqs. 15-17 can be used to calculate the coupling of the molecule with an impinging (light) electric field $\boldsymbol{E}(x,y,z)$ with a sub-nm spatial modulation, as the one in the STM cavity. We can define an effective oscillator strength as:

$$\tilde{f}^j \propto \left| \int_{-\infty}^{\infty} \int_{-\infty}^{\infty} \int_{-\infty}^{\infty} \boldsymbol{E}(x,y,z) \cdot \boldsymbol{\mu}^j(x,y,z) \, dx \, dy \, dz \right|^2. \tag{18}$$

In our analysis, we have considered several functional forms for the electric field in eq.18, which is in the end computed numerically on the spatial grid of the spin orbitals. The results of these tests are omitted for brevity and are available upon reasonable request. Finally, we modelled the electric field in the STM cavity as generated by two point dipoles, as described in the next section.

Two Point-Dipole Modelling of the Electric Field in the STM nanocavity

A simple yet effective model for the electric field in the STM cavity assumes it is generated by two point dipoles [8]: one located at the STM tip and the other at its image position on the metal surface. Both dipoles are assumed to be aligned in the same direction, perpendicular to the xy-plane, which is parallel to the supporting surface and the molecular plane. In these conditions, the components of the electric field are given by:

$$E^x(x,y,z) = \left[ \frac{3p(x-x_T)(z-z_T)}{((x-x_T)^2+(y-y_T)^2+(z-z_T)^2)^{5/2}} + \frac{3p(x-x_T)(z-z_S)}{((x-x_T)^2+(y-y_T)^2+(z-z_S)^2)^{5/2}} \right], \tag{19}$$

$$E^y(x,y,z) = \left[ \frac{3p(y-y_T)(z-z_T)}{((x-x_T)^2+(y-y_T)^2+(z-z_T)^2)^{5/2}} + \frac{3p(y-y_T)(z-z_S)}{((x-x_T)^2+(y-y_T)^2+(z-z_S)^2)^{5/2}} \right], \tag{20}$$

$$E^z(x,y,z) = \left[ \frac{3p(z-z_T)^2}{((x-x_T)^2+(y-y_T)^2+(z-z_T)^2)^{5/2}} - \frac{p}{((x-x_T)^2+(y-y_T)^2+(z-z_T)^2)^{\frac{3}{2}}} + \frac{3p(x-x_T)(z-z_S)}{((x-x_T)^2+(y-y_T)^2+(z-z_S)^2)^{\frac{5}{2}}} - \frac{p}{((x-x_T)^2+(y-y_T)^2+(z-z_S)^2)^{3/2}} \right]. \tag{21}$$

In eqs.19-21 $p$ represents the intensity of the dipole ($p$ is a common term and accounts for a multiplicative factor only), $x_T$ $y_T$, and $z_T$ are the cartesian components of the tip position, whereas $z_S$ is the vertical position of the point dipole in the metal surface (notice that we are assuming that the in-plane position of the point dipole in the metal surface is equal to the ones of the tip, i. e. $x_S = x_T$ and $y_S = y_T$). Since the surface point dipole is located at the image position of the tip inside the surface, $z_S$ is in principle determined, once $z_T$ is set, knowing the number of NaCl layers deposited on the surface. In this work we have set the origin of the z axis on the molecular plane and used $z_T = 11$ au and $z_S = -56$ au, which account for a total of 3 NaCl layer on the metallic surface, and are consistent with the values used in the work by Jaculbia et al. [8]. We have carried out several tests, varying these input values and finding qualitatively similar results as the ones given in the main text, provided that the vertical tip position was set to $z_T \geq 10$ au. We conclude this section observing that the intensity of the electric field, as given in eqs.19-21 diverges at $\{x, y, z\} = \{x_T, y_T, z_T\}$ and $\{x, y, z\} = \{x_T, y_T, z_S\}$. The latter point, being very deep in the metallic surace, is never included in the spatial grid used to integrate eq. 18. As for the divergence in $\{x_T, y_T, z_T\}$, we have verified that all the spin orbital used for the computations in eqs. 15-18 vanish for $z > 10$ au.